\documentclass[12pt]{iopart}

\usepackage{graphicx}
\usepackage[figuresright]{rotating}
\begin{document}

\title[Heavy Flavor Production at STAR]{Heavy Flavor Production at STAR}

\author{Haibin Zhang (for the STAR\footnote[1]{For the full list of STAR authors and acknowledgements, see appendix `Collaboration' of this volume} Collaboration)}

\address{Physics Department, Brookhaven National Laboratory, Upton, NY, 11973, USA}
\ead{haibin@bnl.gov}
\begin{abstract}
We present measurements on $D^0$ meson production via direct
reconstruction of its hadronic decay channel $D^0\rightarrow K\pi$
in minimum bias $d$+Au and Au+Au collisions at $\sqrt{s_{NN}}$=200
GeV with $p_T$ up to $\sim$3 GeV/$c$. Non-photonic electron spectra
from the charm semi-leptonic decays are analyzed from the same data
set as well as in $p$+$p$ collision at $\sqrt{s}$=200 GeV using the
STAR Time-of-Flight (TOF) and Barrel EMC (BEMC) detectors,
respectively. Results of the charm-decayed single muon (prompt muon)
spectra are also presented at low $p_T$ in Au+Au collisions measured
by the TOF detector. The charm production total cross-section per
nucleon-nucleon collision is measured to be
1.26$\pm$0.09(stat.)$\pm$0.23(sys.) mb in minimum bias Au+Au
collisions, which is consistent with the $N_{bin}$ scaling compared
to 1.4$\pm$0.2$\pm$0.4 mb in minimum bias $d$+Au collisions, and
supports the idea that charm quarks should be produced mostly via
parton fusion at early stage in relativistic heavy-ion collisions. A
Blast-Wave model fit to the low $p_T$ ($<2$ GeV/c) non-photonic
electrons, prompt muons and $D^0$ spectra shows that charm hadrons
may kinetically freeze-out earlier than light hadrons with a smaller
collective velocity. The nuclear modification factors ($R_{AA}$) of
the non-photonic electrons in central Au+Au collisions are
significantly below unity at $p_T>\sim$2 GeV/$c$, which indicates a
significant amount of energy loss for heavy quarks in Au+Au
collisions. The charm transverse momentum distribution must have
been modified by the hot and dense matter created in central Au+Au
collisions at RHIC.
\end{abstract}

\pacs{25.75.Dw, 13.20.Fc, 13.25.Ft, 24.85.+p}

\section{Introduction}

Recent experimental studies at the Relativistic Heavy-Ion Collider
(RHIC) have given strong evidences that the nuclear matter created
in Au+Au collisions at $\sqrt{s_{NN}}$=200 GeV has surprisingly
large collectivity and opacity as reflected by its hydrodynamic
behavior at low $p_T$~\cite{flow} and its particle suppression
behavior at high $p_T$~\cite{highpt}. This has led to the famous
name of this high energy density and high temperature nuclear
matter, sQGP, which can be interpreted as either
strongly-interacting Quark Gluon Plasma~\cite{s1}, or
super/perfect-fluid Quark Gluon Plasma~\cite{s2}. However, many of
its important properties are still remain unclear so far, such as
whether the newly-created partonic matter has been thermalized or
not, etc.

Charm quarks can provide a unique tool to probe the partonic matter
created in relativistic heavy-ion collisions at RHIC energies.
First, charm quarks are produced in the early stages of high-energy
heavy-ion collisions due to its relatively large mass~\cite{lin}.
Thus the charm total cross-section is believed to follow the
$N_{bin}$ scaling from $p$+$p$, $d$+Au collisions to Au+Au
collisions at RHIC energies if the nuclear modification to the
parton structure function, the so-called EMC effect~\cite{effect},
is small. The direct measurement of $D^0$ mesons with low $p_T$
coverage in Au+Au collisions will allow us to extract this important
information on the scaling properties of the charm production
cross-section by comparing with the same measurement in $d$+Au
collisions. Second, theoretical calculations have shown that the
charm quarks interacting with the surrounding partons in the medium
could change its flow properties~\cite{teaney,rapp}, such as its
$p_T$ spectra shape, and could boost the elliptic flow ($v_2$) of
the final observable charmed hadrons besides the $v_2$ effect picked
up by their light constituent quarks. Thus experimental measurements
for the $p_T$ spectra of the charmed hadrons and/or its decayed
non-photonic electrons/positrons together with their elliptic flow
properties in Au+Au collisions are particularly interesting to
interpret the thermalization processes of the light quarks in the
partonic matter. Third, charm quarks are believed to lose much
smaller energies compared to light quarks in the partonic matter due
to the famous so-called ``dead-cone"
effect~\cite{dead,miko,armesto}. A measurement of the nuclear
modification factor for the charmed hadrons and/or their decayed
non-photonic electrons/positrons compared to light hadrons is
valuably important to complete the picture of the observed
jet-quenching phenomenon and help us better understand the
energy-loss mechanisms at parton stage in Au+Au collisions at RHIC.

\section{Analysis}

The data used for this analysis were taken with the STAR experiment
during the $\sqrt{s_{NN}}$=200 GeV Au+Au run in 2004 and the
$\sqrt{s_{NN}}$=200 GeV $d$+Au and $p$+$p$ run in 2003 at RHIC. A
minimum bias (minbias) Au+Au collision trigger was defined by
requiring coincidences between two zero degree calorimeters (ZDCs).
A 0-12\% central Au+Au collision trigger was defined using the
scintillator CTB (Central Trigger Barrel) and both the ZDCs. A 0-5\%
central data set is further selected by cutting on the event
multiplicity in the 0-12\% central data sample. The minimum bias
event sample is subdivided into two centrality bins: 10-40\% and
40-80\% for the BEMC non-photonic electron analysis. A minbias
$d$+Au collision trigger was defined by requiring at least one
spectator neutron in the outgoing Au beam direction depositing
energy in a ZDC. A minbias $p$+$p$ collision trigger was defined by
coincidences between two BBCs (Beam-Beam Counter).

\begin{figure}[htp]
\centering
\includegraphics[height=17pc,width=30pc]{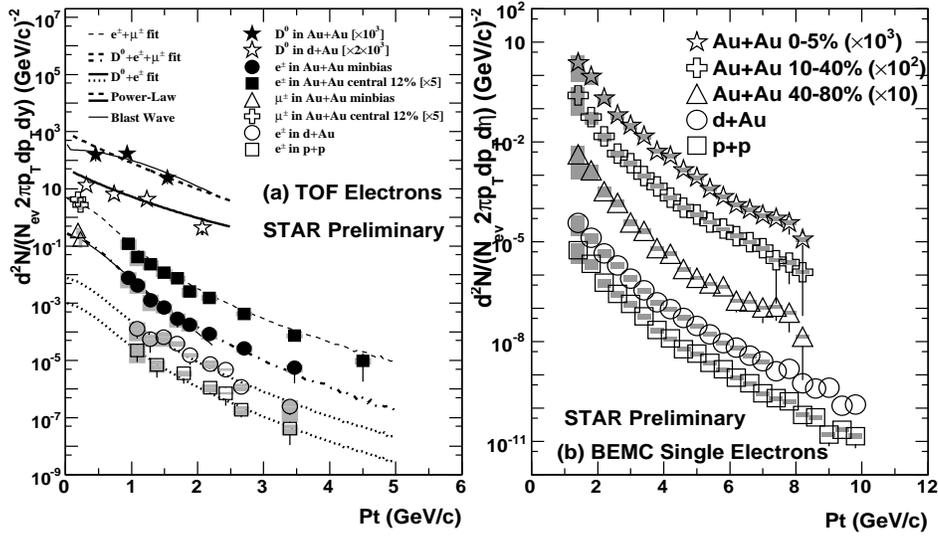}\centering
\caption{(a) $p_T$ distributions of invariant yields for $D^0$
mesons in minbias Au+Au (sold stars) and $d$+Au (open stars)
collisions, charm-decayed prompt muons in 0-12\% central Au+Au (open
crosses) and minbias Au+Au (open triangles) collisions and
non-photonic electrons in 0-12\% central Au+Au (solid squares),
minbias Au+Au (solid circles), $d$+Au (open circles) and $p$+$p$
(open squares) collisions measured by the TOF detector. (b) $p_T$
distributions of invariant yields for single electrons in 0-5\%
central (stars), 10-40\% (crosses), 40-80\% (triangles) Au+Au,
$d$+Au (circles) and $p$+$p$ (squares) collisions measured by the
BEMC detector.}\label{fig:figure1}
\end{figure}

The low $p_T$ ($<3$ GeV/$c$) $D^0$ mesons were reconstructed in
minbias Au+Au and $d$+Au collisions through their decay
$D^0\rightarrow K^-\pi^+$ ($\bar{D^0}\rightarrow K^+\pi^-$) with a
branching ratio of 3.83\%. Analysis details can be found in
Ref.~\cite{dAuCharm,haibin}. Panel (a) of Fig.~\ref{fig:figure1}
shows the $p_T$ distributions of invariant yields for the $D^0$
mesons in minbias Au+Au (solid stars) and $d$+Au (open stars)
collisions.

The charm-decayed prompt muons ($\mu^{\pm}$) at 0.17$<$$p_T$$<$0.25
GeV/$c$ were analyzed by combining the energy-loss information
measured by the STAR Time Projection Chamber (TPC) and the
mass-square ($m^2$) information measured by the TOF detector.
Analysis details can be found in Ref.~\cite{yifei}. The $p_T$
distribution for $\mu^{\pm}$ invariant yields in 0-12\% central and
minbias Au+Au collisions is shown in panel (a) of
Fig.~\ref{fig:figure1}.

By using the combined information from the STAR TPC and TOF
detectors, electrons can be identified and measured. Detailed
analysis for the inclusive, photonic and non-photonic electron
reconstruction can be found in Ref.~\cite{dAuCharm}. The $p_T$
spectra for non-photonic electrons measured by TOF in 0-12\% central
Au+Au (solid squares), minbias Au+Au (solid circles), $d$+Au (open
circles) and $p$+$p$ (open squares) collisions are shown in panel
(a) of Fig.~\ref{fig:figure1}.

Electrons can also be identified by using the STAR TPC and BEMC
detectors as shown in Ref.~\cite{jaro,heavy}. Panel (b) of
Fig.~\ref{fig:figure1} shows the non-photonic electron spectra
measured by BEMC for 0-5\% central (stars), 10-40\% (pink crosses),
40-80\% (triangles) Au+Au, $d$+Au (circles) and $p$+$p$ (squares)
collisions.

\section{Results}

\begin{figure}[htp] \centering
\includegraphics[height=17pc,width=30pc]{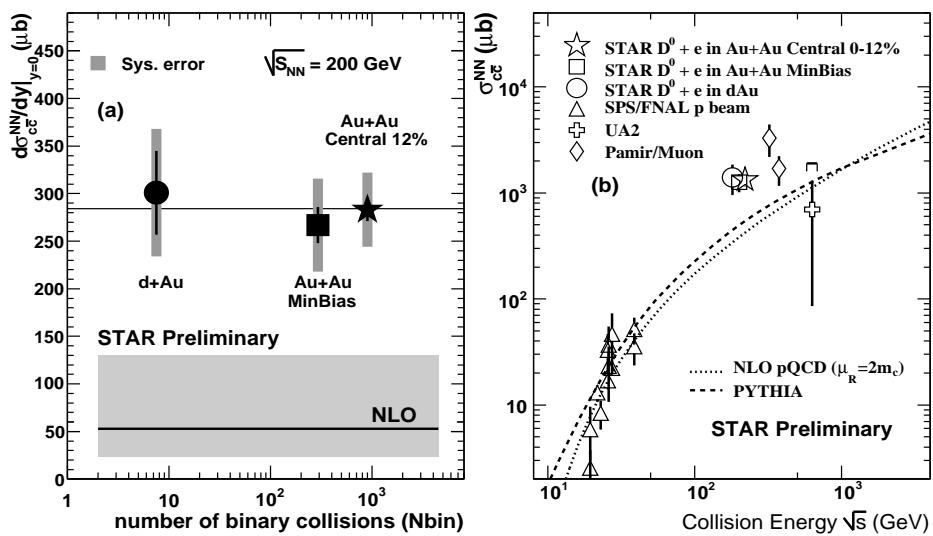}\centering
\caption{(a) Mid-rapidity charm total cross-section per
nucleon-nucleon collision as a function of number of binary
collisions ($N_{bin}$) in $d$+Au, minbias and 0-12\% central Au+Au
collisions. (b) Charm total cross-section per nucleon-nucleon
collision as a function of collision energy ($\sqrt{s}$) in $d$+Au,
minbias and 0-12\% central Au+Au collisions compared various
collision systems with various collision
energies.}\label{fig:figure2}
\end{figure}

Using a combined fit applied to the directly reconstructed $D^0$
spectra, charm-decayed prompt muon spectra and the non-photonic
electron spectra in 0-12\% central Au+Au, minbias Au+Au and $d$+Au
collisions, the mid-rapidity $D^0$ yield is then obtained and
converted to the mid-rapidity charm total cross-section per
nucleon-nucleon collision ($d\sigma_{c\bar{c}}^{NN}/dy$) and the
charm total cross-section per nucleon-nucleon collision
($\sigma_{c\bar{c}}^{NN}$) following the method addressed in
Ref.~\cite{dAuCharm}. $\sigma_{c\bar{c}}^{NN}$ is measured to be
1.33$\pm$0.06(stat.)$\pm$0.18(sys.) mb in 0-12\% central Au+Au,
1.26$\pm$0.09$\pm$0.23 mb in minimum bias Au+Au collisions and
1.4$\pm$0.2$\pm$0.2 mb in minimum bias $d$+Au collisions at
$\sqrt{s_{NN}}$=200 GeV. Panel (a) of Fig.~\ref{fig:figure2} shows
the $d\sigma_{c\bar{c}}^{NN}/dy$ as a function of $N_{bin}$ for
minbias $d$+Au, minbias Au+Au and 0-12\% central Au+Au collisions.
It can be observed that the charm total cross-section roughly
follows the $N_{bin}$ scaling from $d$+Au to Au+Au collisions which
supports the conjecture that charm quarks are produced at early
stages in relativistic heavy-ion collisions. Panel (b) of
Fig.~\ref{fig:figure2} shows the $\sigma_{c\bar{c}}^{NN}$ as a
function of $\sqrt{s}$ for minbias $d$+Au, minbias Au+Au and 0-12\%
central Au+Au collisions compared various collision systems at
various collision energies as well as theoretical predictions.
However, one can clearly see from Fig.~\ref{fig:figure2} that the
$d\sigma_{c\bar{c}}^{NN}/dy$ in the three measured collision systems
are about a factor of 5 larger than the NLO
predictions~\cite{vogt1,vogt2} showns as the light green band in
panel (a).

\begin{figure}[htp]
\centering
\includegraphics[height=17pc,width=24pc]{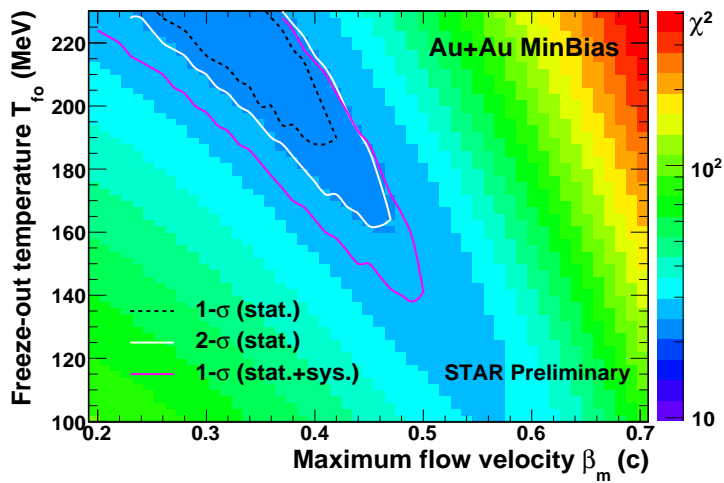}\centering
\caption{(a) Charm hadron freeze-out temperature ($T_{fo}$) versus
maximum flow velocity ($\beta_m$) in minbias Au+Au
collisions.}\label{fig:figure3}
\end{figure}

A Blast-Wave model~\cite{blast} fit to the $D^0$, prompt muons and
non-photonic electron $p_T$ spectra at $p_T<$2 GeV/$c$ in minbias
Au+Au collisions, from which the charm hadron kinetic freeze-out
temperature $T_{fo}$ and the maximum flow velocity $\beta_m$ are
derived. Fig.~\ref{fig:figure3} shows the $T_{fo}$ versus $\beta_m$
for charm hadrons in minbias Au+Au collisions. The 1$\sigma$ contour
from the Blast-Wave fit with quadratic sum of statistical and
systematic errors of the spectra is shown as the pink curve in
Fig.~\ref{fig:figure3}. A larger $T_{fo}$ ($>$140 MeV) and a smaller
$\beta_m$ compared to those of light hadrons~\cite{hadron} can be
observed from the figure, which indicate that charm hadrons may
kinetically freeze-out early and may not be in complete equilibrium
with the rest of the system at kinetic freeze-out in minbias Au+Au
collisions.

\begin{figure}[htp]
\centering
\includegraphics[height=17pc,width=30pc]{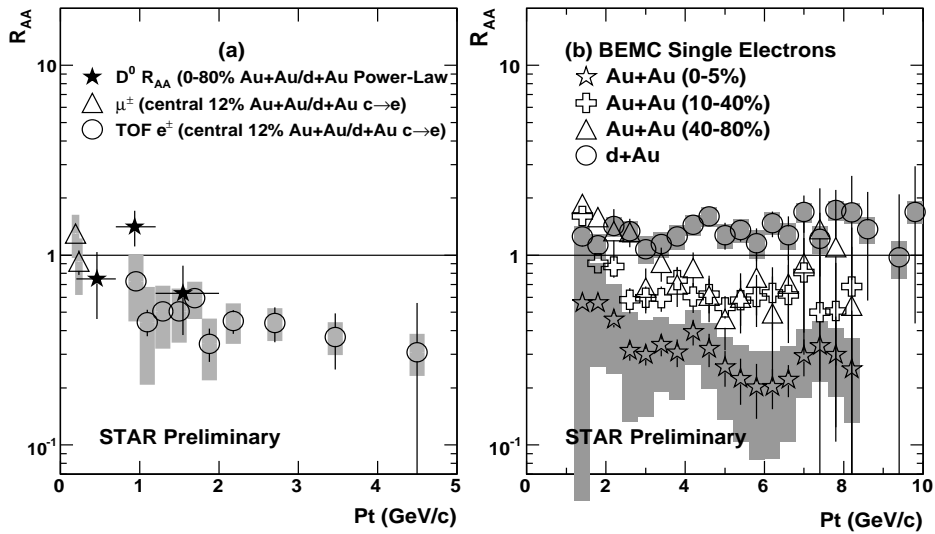}\centering
\caption{(a) $p_T$ distributions of the nuclear modification factor
($R_{AA}$) for $D^0$, charm-decayed prompt muons and single
electrons measured by the TOF detector. (b) $p_T$ distributions of
$R_{AA}$ for non-photonic electrons in 0-5\% central, 10-40\%,
40-80\% Au+Au and $d$+Au collisions measured by the BEMC
detector.}\label{fig:figure4}
\end{figure}

The $D^0$ $R_{AA}$ (stars in Panel (a) of Fig.~\ref{fig:figure4})
are calculated by dividing the $D^0$ data points in minimum bias
Au+Au collisions by the power-law fit results of the $D^0$ $p_T$
spectrum in $d$+Au collisions scaled by $N_{bin}$. The prompt muon
$R_{AA}$ are calculated by dividing the $p_T$ spectrum in 0-12\%
central Au+Au collisions to that in minbias Au+Au collisions with
$N_{bin}$ scaling, shown as triangles in panel (a) of
Fig.~\ref{fig:figure4}. The TOF-measured single electron $R_{AA}$
are also calculated by dividing the $p_T$ spectra in 0-12\% central
Au+Au collisions to the $D^0\rightarrow e^{\pm}$ decayed shape in
$d$+Au collisions scaled by $N_{bin}$, shown as open circles in
panel (a) of Fig.~\ref{fig:figure4}. The $R_{AA}$'s for $D^0$ and
muons at low $p_T$ are consistent with unity considering
uncertainties. The non-photonic electron $R_{AA}$ in 0-12\% central
Au+Au collisions is observed to be significantly below unity at
1$<p_T<$4 GeV/$c$.

The BEMC-measured non-photonic electron $R_{AA}$'s for 0-5\% central
(stars), 10-40\% (crosses), 40-80\% (triangles) Au+Au collisions and
$d$+Au (circles) collisions are calculated by dividing the
corresponding non-photonic electron $p_T$ spectra to that in $p$+$p$
collisions, respectively, and are shown in panel (b) of
Fig.~\ref{fig:figure4}. The non-photonic electron $R_{AA}$ in $d$+Au
collisions ($R_{dAu}$) is close to/above unity indicating the Cronin
effect in $d$+Au collisions. In Au+Au collisions, the $R_{AA}$ in
central collisions is smaller than in peripheral collisions. The
$R_{AA}$ in 0-5\% central Au+Au collisions is significantly below
unity at $p_T>$2 GeV/$c$ and is suppressed as strongly as that of
light hadrons~\cite{highpt}, which indicates a large amount of
energy-loss for heavy quarks in central Au+Au collisions.

According to the famous ``dead-cone"
effect~\cite{dead,miko,armesto}, bottom quarks should lose smaller
energy than charm quarks due to their mass difference. Theoretical
calculations~\cite{armesto,mag} considering only the charm
contributions to the non-photonic electrons agree with the measured
non-photonic electron $R_{AA}$, while calculations with single
electrons decayed from both bottom and charm quarks give larger
$R_{AA}$ values. However, in most theoretical models, the amount of
bottom quark and charm quark contributions to the non-photonic
electron spectra, respectively, still remains uncertain. Thus, an
experimental measurement of the $R_{AA}$'s from directly
reconstructed charm hadrons ($D^0$, $D^{\pm}$, $D_S$, $\Lambda_C$,
etc.) at high $p_T$ is necessary. A detector upgrade plan for a
sillicon pixel detector, the Heavy Flavor Tracker (HFT)~\cite{hft},
at STAR will make it possible for the direct charm hadron $R_{AA}$
measurements in the near future.

\section{Conclusion}

We present measurements on $D^0$ meson production via direct
reconstruction of its hadronic decay channel $D^0\rightarrow K\pi$
in minimum bias $d$+Au and Au+Au collisions at $\sqrt{s_{NN}}$=200
GeV with $p_T$ up to $\sim$3 GeV/$c$. Non-photonic electron spectra
from the charm semi-leptonic decays are analyzed from the same data
set as well as in $p$+$p$ collision at $\sqrt{s}$=200 GeV using the
STAR Time-of-Flight (TOF) and Barrel EMC (BEMC) detectors,
respectively. Results of the charm-decayed prompt muon spectra are
also presented at low $p_T$ in Au+Au collisions measured by the TOF
detector. The charm production total cross-section per
nucleon-nucleon collision is measured to be
1.26$\pm$0.09(stat.)$\pm$0.23(sys.) mb in minimum bias Au+Au
collisions, which is consistent with the $N_{bin}$ scaling compared
to 1.4$\pm$0.2$\pm$0.4 mb in minimum bias $d$+Au collisions, and
supports the idea that charm quarks should be produced mostly via
parton fusion at early stage in relativistic heavy-ion collisions. A
Blast-Wave model fit to the low $p_T$ ($<2$ GeV/$c$) non-photonic
electrons, prompt muons and $D^0$ spectra shows that charm hadrons
may kinetically freeze-out earlier than light hadrons with a smaller
collective velocity. The nuclear modification factors ($R_{AA}$) of
the non-photonic electrons in central Au+Au collisions are
significantly below unity at $p_T>\sim$2 GeV/$c$, which indicates a
significant amount of energy loss for heavy quarks in Au+Au
collisions. The charm transverse momentum distribution must have
been modified by the hot and dense matter created in central Au+Au
collisions at RHIC.

\end{document}